# Intrinsic limits governing MBE growth of Ga-assisted GaAs nanowires on Si(111)

Le Thuy Thanh Giang, C. Bougerol, H. Mariette, and R. Songmuang[*]

*CEA-CNRS Group "Nanophysique and Semiconducteurs, Institute Néel, 17 Rue des Martyrs, 38054 Grenoble cedex 9, France*

Diffusion-enhanced and desorption-limited growth regimes of Ga-assisted GaAs nanowires were identified. In the latter regime, the number of vertical NWs with a narrow length distribution was increased by raising the growth temperature. The maximum axial growth rate; which can be quantified by the supplied rate of As atoms, is achieved when a dynamical equilibrium state is maintained in Ga droplets i.e. the number of impinging As atoms on the droplet surface is equivalent to that of direct deposited Ga atoms combining with the diffusing ones. The contribution of Ga diffusion to the wire growth was evidenced by the diameter-dependent NW axial growth rate.

Corresponding author: rudeesun.songmuang@grenoble.cnrs.fr

Semiconductor nanowires (NWs) have gained an increasing interest because of their various unique features ranging from the fundamental physic of their growth to their electronic/optic/mechanical properties [1]. A widely used technique to synthesize III-As NWs is a vapor-liquid-solid (VLS) process [2], in which an external metal catalyst such as Au is required. Indeed, Au catalysts could be replaced by group III-metal, offering an advantage of the wire growth in Au-free environment [3,4]. In order to enlarge the degree of freedom to control the NW properties, knowledge concerning their growth mechanism is demanding. Indeed, the insight of physical effects (e.g. adsorption, desorption, diffusion) which govern their growth behavior could be accessed by correlating the NW morphology with technological growth parameters.

Recently, molecular beam epitaxy (MBE) growth of Ga-assisted GaAs NWs was demonstrated by several teams on either GaAs [3] or Si substrates [5,6,7,8]. Although a few works presented recently focused on studying their growth behavior [7,9,10,11], thorough investigations concerning the influences of basic growth parameters on the NW growth remains required to get deeper understanding of their growth mechanism. For examples, the effect of the substrate temperatures on the NW morphology was reported [6, 7], but the issues concerning the diffusion and the desorption process which are usually discussed were not experimentally evidenced. Although it was shown that As flux is a controlling parameter for the NW axial growth [9, 10], the maximum limit was not clearly quantified. In addition, despite an intensive discussion on the contribution of Ga and As adatom diffusion to the axial growth [6,7,10], the diameter dependent axial growth rate were not experimentally demonstrated and compared with the theoretical models, unlike the case of Au-catalyzed NW growth.

In order to complete the knowledge available in literatures, we perform detailed and systematically analyses of Ga-assisted NW structural evolution as a function of MBE growth parameters such as substrate temperature, III/As ratio and deposition rate. In our experiments, the supplied rates of As atoms were measured, thus the number of impinging As atoms and As/Ga ratio can be quantitatively determined. In addition, the axial growth rates were statistically analyzed as a function of the wire diameter. The experimental data were fitted with the theoretical model [12] in order to deduce the diffusion length which is an important characteristic of their growth kinetics. Finally, we show that the NW growth can approach a near-equilibrium condition by simply fine tuning the basic technological parameters.

Ga-assisted GaAs NWs were grown by using solid source MBE Riber32P equipped with a conventional $As_4$ source. The substrates are Si(111) covered with native oxide layer. Prior to the NW growth, the substrates were annealed at 690°C for 10 min and then Ga and As fluxes were simultaneously deposited on the surface. The relatively low *in-situ* annealing temperature was intentionally chosen to create a discontinuity in the oxide layers rather than completely desorb them from the surface [13,14]. The nominal Ga deposition rate was measured on separated GaAs (001) substrates via reflection high-energy electron diffraction (RHEED) intensity oscillations. The deposition rate of As atoms were approximately similar to that of Ga atoms which induces the RHEED pattern transition from (2x4) As stabilized to (4x2) Ga stabilized surface [15,16]. This calibration measures the As atoms which are actually incorporated into the growing GaAs. The NW structural properties were probed by using field emission scanning electron microscope (FESEM) and transmission electron microscope (TEM).

The selected side view SEM images shown in Figs. 1(a)-(b) present GaAs NWs grown at 560°C and 600°C, respectively. In this series, the nominal deposition rates of Ga and As were fixed at 0.1 ML/s and 0.2 ML/s, respectively while the deposition time was kept at 45 min. The NWs were separated into 2 categories i.e. the non-vertical [Fig. 1(a)] and the vertical NWs [Fig.

1(b)]. At growth temperature above 600°C, the NWs usually show invert-tapering shape, implying the expansion of Ga droplets during the growth. This is attributed to a strong degree of As desorption, inducing a lack of available As atoms for the arriving Ga ones. Although the growth temperature strongly influences on the NW morphology, it has a negligible effect on their crystallographic structure as mentioned in ref. [7]. Our TEM images show that the NW adopt mainly zinc blend (ZB) structure with a few twin plane and stacking faults in the whole range of studied temperatures.

Figure 1(c) presents the density evolution of the non-vertical and vertical wires summarized as a function of substrate temperature. The total density of the NWs including Ga droplets on the surface decreases from $10^9$ to $10^8$ cm$^{-2}$ (not shown here), following the same tendency as that of the non-vertical NWs. This evolution suggests a pronounce material desorption at high growth temperature. At around 600°C, the vertical wires become observable and their density increases with the substrate temperature. Because GaAs NWs grow along [111]B direction [17], the NWs perpendicular to the substrate imply an As terminated (111) surface below Ga droplets; that is, As atoms substitute the outermost layer of Si(111) surface [18]. From the increasing density of the vertical wires as a function of substrate temperature, we deduce that the formation of such an interface is enhanced by using high growth temperature.

The average diameter and length of the vertical and the non-vertical wires were plotted as a function of the substrate temperature shown in Figs. 2(a)-(b). These values of both types of NWs increase to the maximum, when the substrate temperature was raised up to 600-620°C and then they tend to decreases. At the growth temperature higher than 700°C, the material growth was not possible at our applied deposition rate. This structural evolution evidences both the *thermally* *enhanced and limited* growth regimes of the NWs [19,20], ascribed to the effects of the diffusion and desorption process, respectively.

During the nucleation stage, the impinging Ga atoms perform a thermal activated random walk on the substrate surface. The collisions between those diffusing adatoms lead to the formation of nuclei of Ga droplets. Below 620°C, where the desorption process is insignificant, raising the growth temperature makes the diffusing adatoms preferentially attaching to the existing droplets rather than nucleating the new ones. Such a surface kinetic could enlarge the droplet size which determines the wire diameter [Fig. 2(a)]. At the growth temperature above 620°C, the dominant desorption process reduces the residence time of adatoms on the surface leading to a significant material loss and a shorter effective diffusion length. This effect explains a strong reduction of the droplet density [Fig. 1(c)] and a smaller diameter of the droplets which later activate the NW growth [Fig. 2(a)].

After the Ga droplets reach a supersaturation condition, the NW growth proceeds by consuming the material directly deposited on the wire top and the one diffusing through the NW sidewall to their apex [21,22,23,24]. The length increment with the growth temperature below 620°C [Fig 2(b)] reveals more material participating in the axial growth induced by a thermally enhanced diffusion. Above 620°C, the monotonically decrease of the wire length is attributed to the large degree of material desorption, which causes less Ga and As atoms contributing to the growth.

Here, the maximum NW axial growth rate is around 8 times faster than the nominal As deposition rate [25]. By considering only a simple geometric effect assuming a spherical shape droplet and a cylindrical shape wire, this obtained factor is consistent with the ratio between the surface area of the Ga droplet exposed to the impinging atoms and the cross-section area of the growing NWs below the droplets. This geometric ratio is around 8 approximated by using the contact angle of Ga droplets equal to 140° measured from SEM images. Such an agreement quantitatively reveals that most of As atoms impinging on Ga droplet surface contribute to the NW axial growth while the As diffusion is negligible. Besides quantitatively confirming that As flux is the limiting parameter [9], the maximum possible axial growth rate can be determined *a priori* from the supplied rate of As atoms.

Clearly, the vertical aligned wires have smaller diameter, longer length and narrower length distribution in comparison to the tilted ones grown under the same conditions [Figs. 2 (a)-(b)]. These differences point-out the effect of the wire axis in respect to the direction of the deposited beam flux [26]. The inclination of NW axis toward the substrate surface increases a probability of the sidewall deposition, leading to higher adatom concentration on this area. This fact possibly reduces the driving force for the material transport toward the wire top, thus suppressing the growth in the axial direction while enhancing the growth in the lateral one. Very high length uniformity of the vertical wires shows that their axial growth rate reaches the maximum limit set by the impinging As atoms on the droplet surface.

In order to quantify the corresponding effective diffusion lengths which represent the NW growth kinetics, the diameter-dependence axial growth rates were investigated. In this case, only Ga adatom diffusion is considered since the As diffusion is negligible. We applied a stationary growth model suggested by Fröberg et al, to describe the axial growth rate ($R_{Axial}=dL/dt$) of a cylindrical NW with a radius ($r_w$) by a competition between Gibb-Thomson effect and surface diffusion process using a following equation [13]:

$$\frac{dL}{dt} = R_{dep}\lambda_s[1-\gamma(r_w)]\frac{2\Omega_{GaAs}}{r_w}\frac{K_1(r_w/\lambda_s)}{K_0(r_w/\lambda_s)} + R_{dep}\Omega_{GaAs} \qquad (1)$$

where $R_{dep}$ is the arrival rate of adatom on substrate surface, $\lambda_s$ is the Ga surface diffusion length, $\Omega_{GaAs}$ is an atomic volume of ZB GaAs, $\gamma(r_w)$ is the dimensionless parameter representing the Gibb-Thomson effect, and $K_0$ and $K_1$ are the modified Bessel function of the second kind of the zero and first order, respectively, while $\gamma(r_w)=(2\sigma\Omega_{Ga})/(r_w\Delta\mu_\infty)$, where $\Delta\mu_\infty$ represents a supersaturation in Ga droplet with an infinite radius of curvature, $\sigma$ is a surface energy of Ga liquid and $\Omega_{Ga}$ is an atomic volume of Ga liquid.

In order to compare the diffusion length, $\lambda_s$, corresponding to the NW growth kinetics in the whole temperature range, we consider the data taken from the tilted NWs which are always found on the substrate surface. As the sidewall deposition contribution is neglected in this model, the extracted diffusion lengths give an *effective* value which can be relatively compared between each growth temperature. Figure 3 shows the relation between the axial growth rate and the wire diameter of the tilted NWs grown at 480°C and 600°C including that of the vertical ones obtained at 600°C. We found that the axial growth rate shows one maximum at a certain diameter range, similar to what observed in Au induced NWs [27]. Although the model is phenomenological, it could describe our data fairly well.

The inset of Fig. 3 is the extracted $\lambda_s$ summarized as a function of the growth temperature. Below 620°C, the increasing $\lambda_s$ from 40 nm to 120 nm is consistent with the thermally enhanced diffusion regime. Above 620°C, $\lambda_s$ decreases downward to 60 nm at 670 °C representing the desorption-limited diffusion. However, at high growth temperature, the extracted $\lambda_s$ might be underestimated since the growth process is limited by a strong As desorption which is actually seen by the invert tapering NW shape.

To better understand the NW growth kinetics, different supersaturation conditions were used for growing the NWs. In this series, the Ga deposition rate was varied from 0.1 to 0.7 ML/s while the As deposition rate was kept at 0.48 ML/s. The substrate temperature and the deposition time were fixed at 640°C and 45 min. Figures 4(a)-(c) present the side view SEM images of the NWs grown by using the Ga deposition rate of 0.1 ML/s, 0.2 ML/s and 0.7 ML/s, respectively. Clearly, the non-monotonic variation of the NW length was observed as a function of As/Ga ratio. Very thin and short GaAs wires appear at the lowest supplied Ga flux or the highest As/Ga ratio, ascribed to a small critical nucleation size induced by high supersaturation condition [Fig. 4(a)]. At As/Ga ratio around 1.5-2.4, the longest NWs were obtained with their axial growth rate roughly 8 times the supplied rate of As atoms [Fig. 4(b)]. When As/Ga ratio was less than 1.5, the larger wire diameter and the important invert tapering degree were observed while their axial growth rate slowly decreases to roughly 3 times of the supplied rate of As at As/Ga ratio ~ 0.7 [Fig. 4(c)].

For large As/Ga ratio (~5), Ga droplets remain observable on the top of the NWs grown [Fig. 4(a)], evidencing that the VLS mechanism remains governing the NW growth. Their low axial growth rate is thus due to an insufficiency of Ga atoms arriving to the droplets, in comparison to the impinging As atoms on the droplet surface. In fact, further depositing GaAs at this As/Ga ratio would result in a complete consummation of Ga droplets on the NW top and then the termination of VLS mechanism. Afterward, the axial growth would drastically decrease and the NW growth would approach two dimensional manners.

On the other hand, the excess of supplied Ga atoms in Ga-rich condition also causes the decrement of the NW axial growth [Fig. 4(c)]. Since Ga atoms were sufficiently supplied in this case, the lower axial growth rate than the maximum possible value indicates less As atoms contributing to the axial growth i.e. the excessively fed Ga atoms to the droplets prevent the droplets to reach a supersaturation. Therefore, the supplied As atoms remain in the liquid alloy rather than participate in the crystal growth in order to maintain a stable As concentration in the expanding Ga droplets at this growth temperature. Further depositing GaAs at this ratio leads to a continuous accumulation of Ga atoms in the droplets, inducing the invert-tapering shape of NWs and possibly a decrease of the axial growth rate while the NWs proceed.

Figure 4(d) shows the axial growth rate plotted as a function of the diameter of the NWs grown at the Ga deposition rate of 0.1, 0.2 and 0.7 ML/s. The solid lines are the curves obtained by fitting the data using eq.1. The shorter effective diffusion length i. e. $\lambda_s$= 58 nm and $\lambda_s$= 36 nm for the Ga deposition rate of 0.1 and 0.7 ML/s, respectively, implies less diffusion contribution to the axial growth caused either by the deficient or excessive Ga atoms arriving to the droplets. We deduce that the highest contribution of the diffusion to the NW growth occurs at As/Ga ratio around 2.4 ($\lambda_s$= 219 nm), where the maximum axial growth rate is obtained. This $\lambda_s$ is higher than the previous case since the data from the vertical aligned wires was taken into account.

Our results show that the maximum axial growth rate can be achieved when a dynamical equilibrium state in the droplet is maintained i.e. the species incorporating to the droplets equal to the ones consuming by the NW growth. This situation occurs when the number of impinging As atoms on the droplet surface is equivalent to the arriving Ga atoms from direct deposition and the diffusion. Hence, As/Ga ratio larger than one on the droplet surface is initially required to approach equilibrium in the droplet. However, if this ratio was kept constant during the growth, the steady state would remain for a certain time as the diffusing material decreases as a function of the wire length.

We further our studies by varying the material deposition speeds while trying to maintain a steady state in the droplets by keeping the As/Ga ~2 at the growth temperatures at 640°C. The nominal deposited GaAs is fixed at 1080 ML. Figures 5(a)-(b) are the side view SEM images of the NWs grown by using the Ga deposition rate of 0.2 ML/s and 0.6 ML/s, respectively. Most of NWs show a non-tapering shape and the NW morphology of each sample is similar although the deposition rate was significantly varied. At our chosen conditions, the NW density and diameter are weakly dependent on the growth speed. Therefore, after the nucleation stage, the NW growth and their morphology are mainly controlled by the material distribution among each nucleation site. Figure 5(c) shows that their axial growth rate is a linear function of the nominal deposition rate. The NW axial growth rate is again around 8-10 times faster than the supplied rate of As, consistent with the maximum limit derived from our geometric approximation described previously. Our results show that once a dynamical equilibrium state is approached, the NW morphology is stable to kinetic parameters.

In conclusion, we present comprehensive studies of the morphology evolution of Ga-assisted GaAs NWs on Si(111) substrates as a function of MBE growth parameters. By raising the substrate temperature, the diffusion-enhanced and desorption-limited growth regimes were identified and the improved vertical-alignment was observed. The contribution of Ga diffusion to the NW growth was evidenced by the diameter-dependent axial growth. Slightly As-rich condition on the droplet surface is needed to approach a steady state in Ga droplets where the axial growth rate reaches the maximum. This growth condition also makes the NW morphology insensitive to the growth speed.

L. T. T. G. acknowledges the scholarship from the ministry of education, Vietnam. The authors thank Y. Genuist and Y. Cure for their technical supports and Le Si Dang for fruitful discussions.

**Figure captions**

**Fig. 1** Side view SEM images of Ga-assisted GaAs NWs grown at (a) 560°C and (b) 600°C. (c) Density of vertical (■) and non-vertical NWs (▲) plotted as a function of substrate temperature

**Fig. 2** (a) Average diameter and (b) length of non-vertical (■) and vertical wires (▲) plotted as a function of substrate temperatures

**Fig. 3** Length and diameter relation of GaAs NWs grown at 480°C (▼)and 600°C (■). The blue and red dash lines are the fit to the average data points using $\lambda_s$= 60 nm, $\Delta\mu_\infty$= 0.21 $kT$, and $\lambda_s$= 112 nm, $\Delta\mu_\infty$=0.12 kT for 480°C and 600°C, respectively, where $k$ is Boltzmann constant and $T$ is substrate temperature. The inset summarized the extracted effective surface diffusion length, $\lambda_s$, as a function of substrate temperature. The grey dotted line is for eye guiding.

**Fig. 4** Side view SEM images of GaAs NWs grown by using Ga deposition rate equal to (a) 0.1 ML/s, (b) 0.2 ML/s and (c) 0.7 ML/s while fixing the As deposition rate at 0.48 ML/s and (d) their length and diameter relation. The solid lines represents the fit to the average data points using $\lambda_s$= 58 nm ($\Delta\mu_\infty$= 0.29 kT), $\lambda_s$= 219 nm ($\Delta\mu_\infty$= 0.1 kT), and $\lambda_s$= 36 nm ($\Delta\mu_\infty$= 0.07 kT), respectively. Note that the data were taken from both non-vertical and vertical wires since the density ratio of two types of wire are similar in all samples.

**Fig. 5** Side view SEM images of GaAs NWs grown by using Ga deposition rate equal to (a) 0.2 ML/s and (b) 0.6 ML/s while keeping As/Ga ratio ~2, the growth temperature at 640°C and the nominal deposited GaAs at 1080ML (c) Average axial growth rate of GaAs NWs as a function of Ga deposition rate.

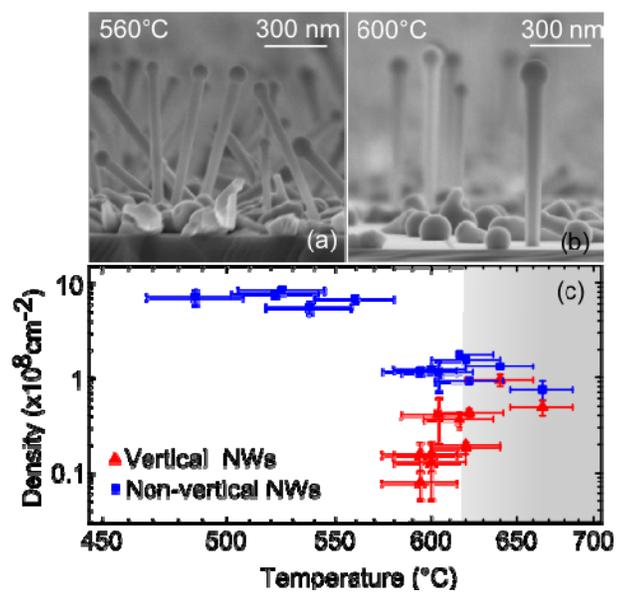

Figure 1

Le Thuy Thanh Giang *et al*

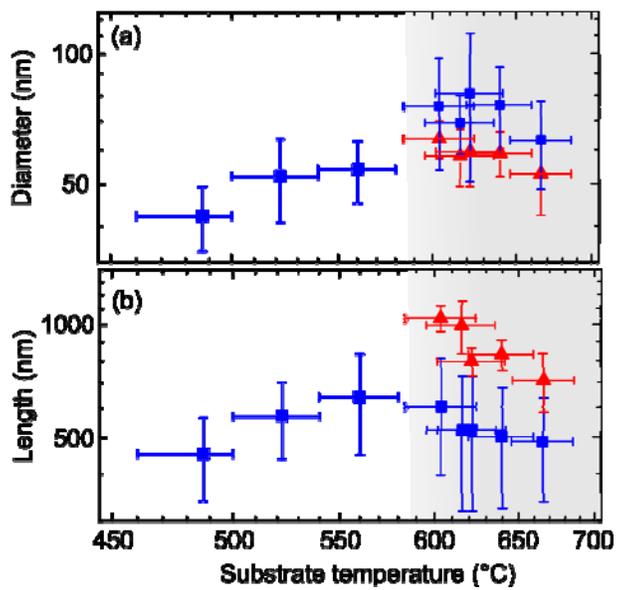

Figure 2

Le Thuy Thanh Giang *et al*

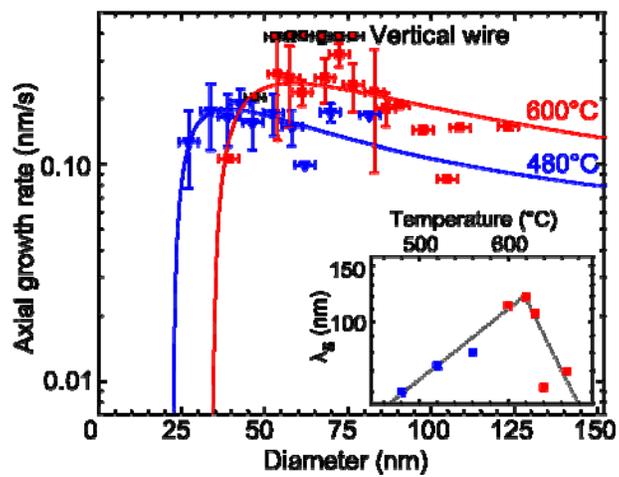

Figure 3

Le Thuy Thanh Giang *et al*

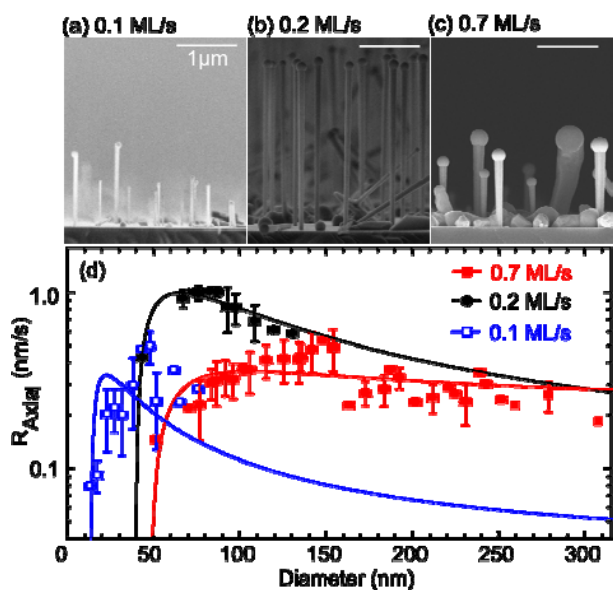

Figure 4

Le Thuy Thanh Giang *et al*

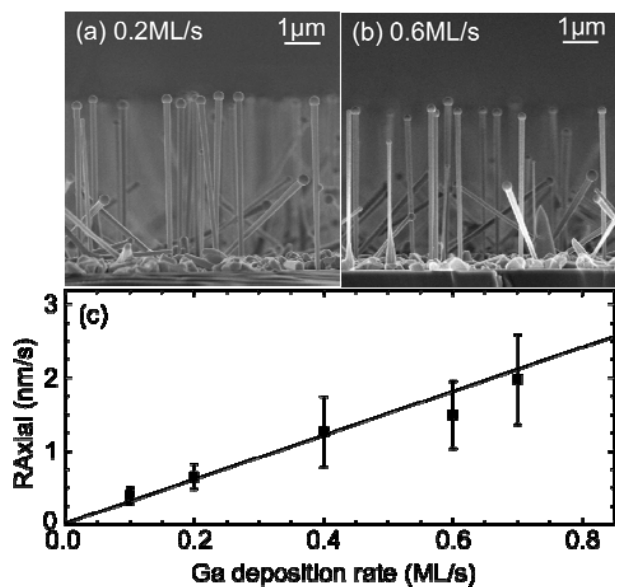

Figure 5

Le Thuy Thanh Giang *et al*

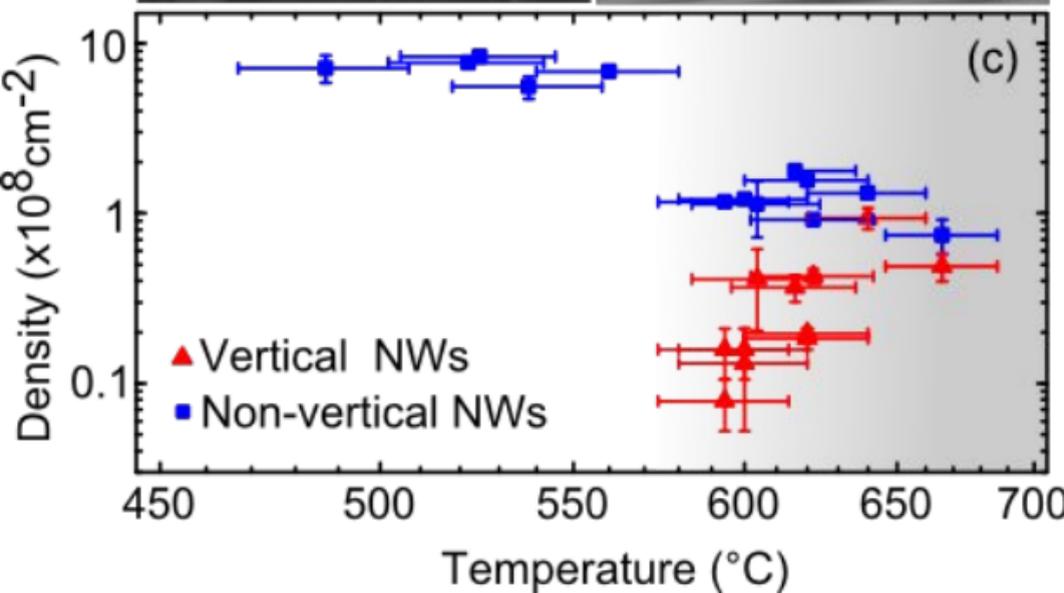

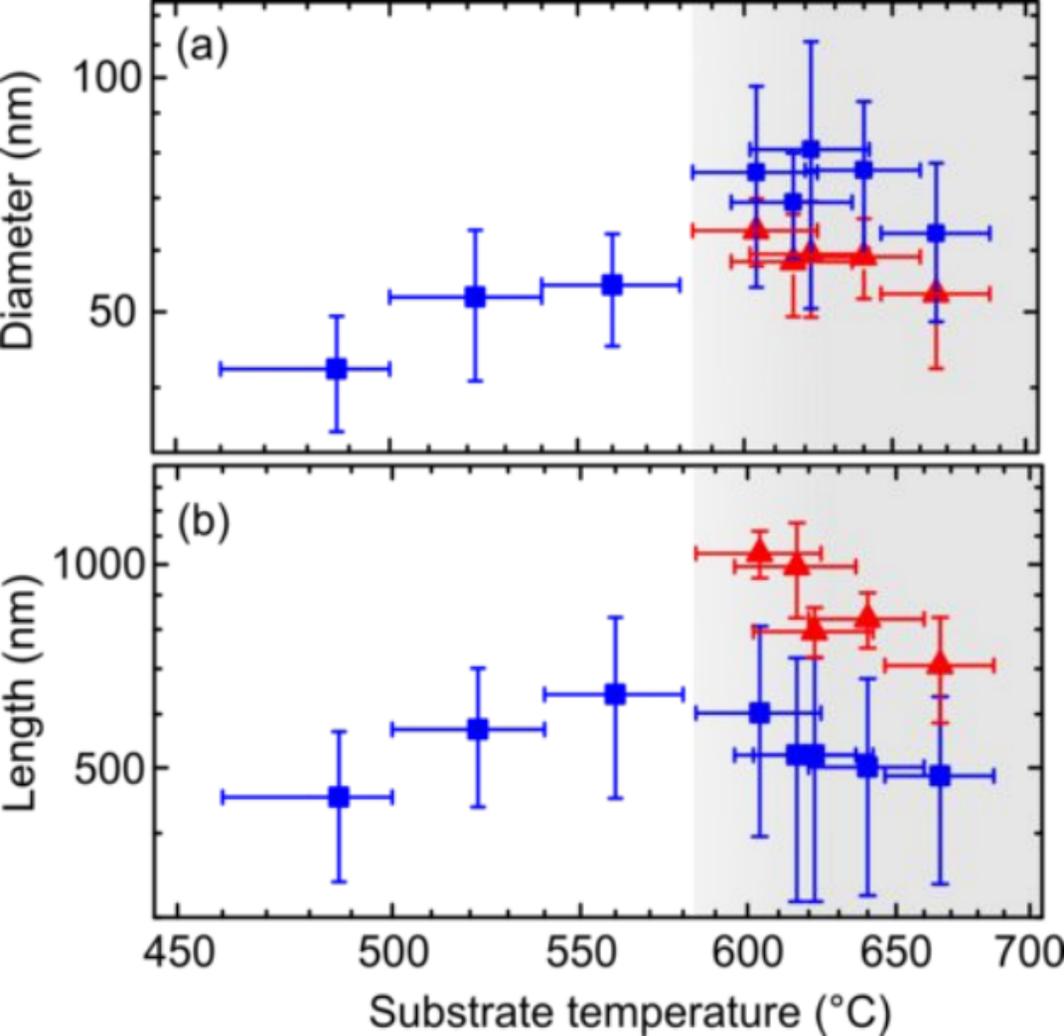

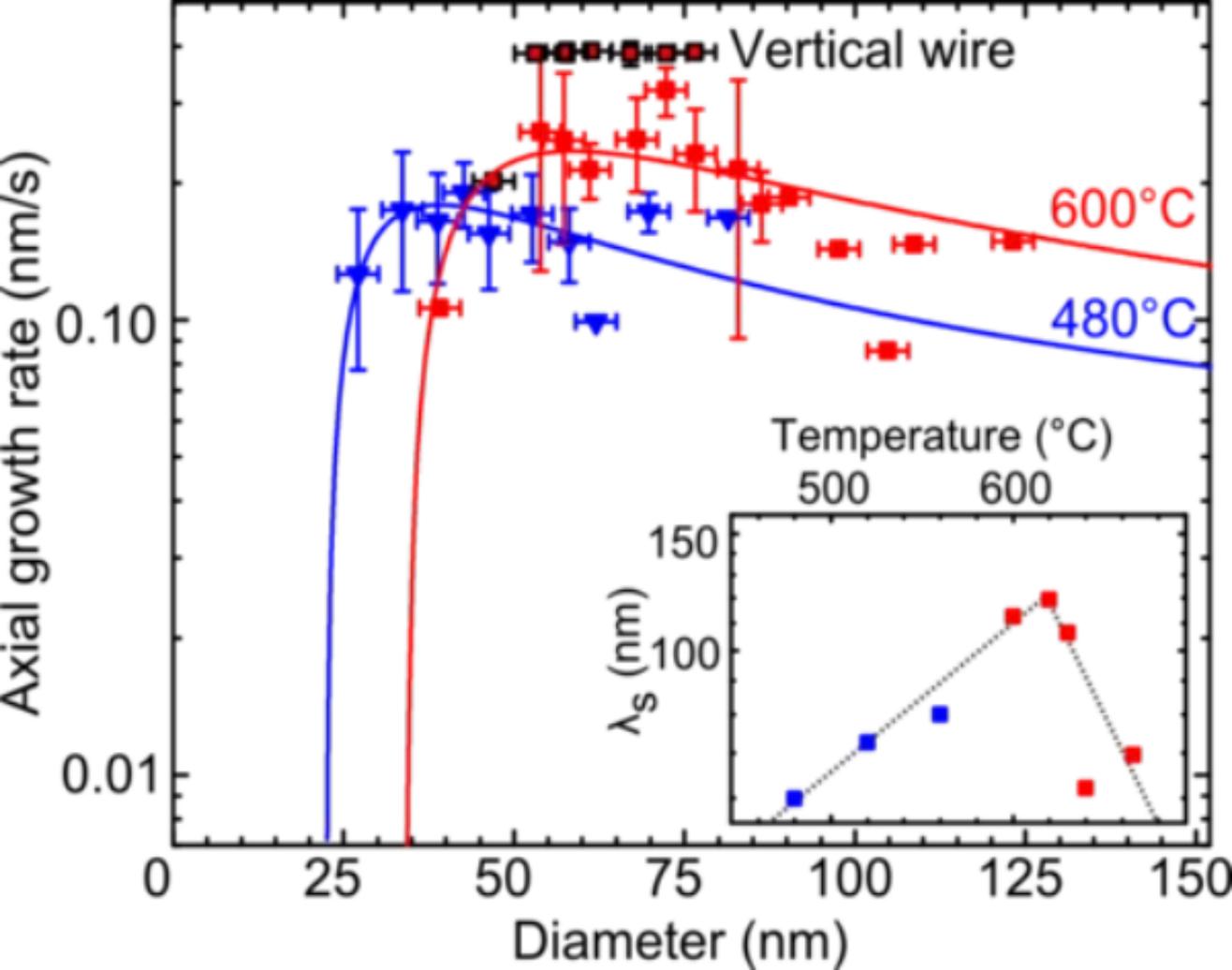

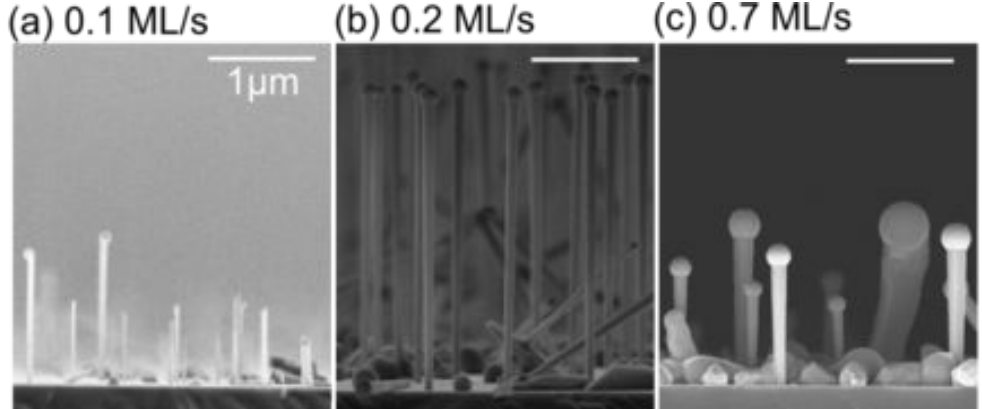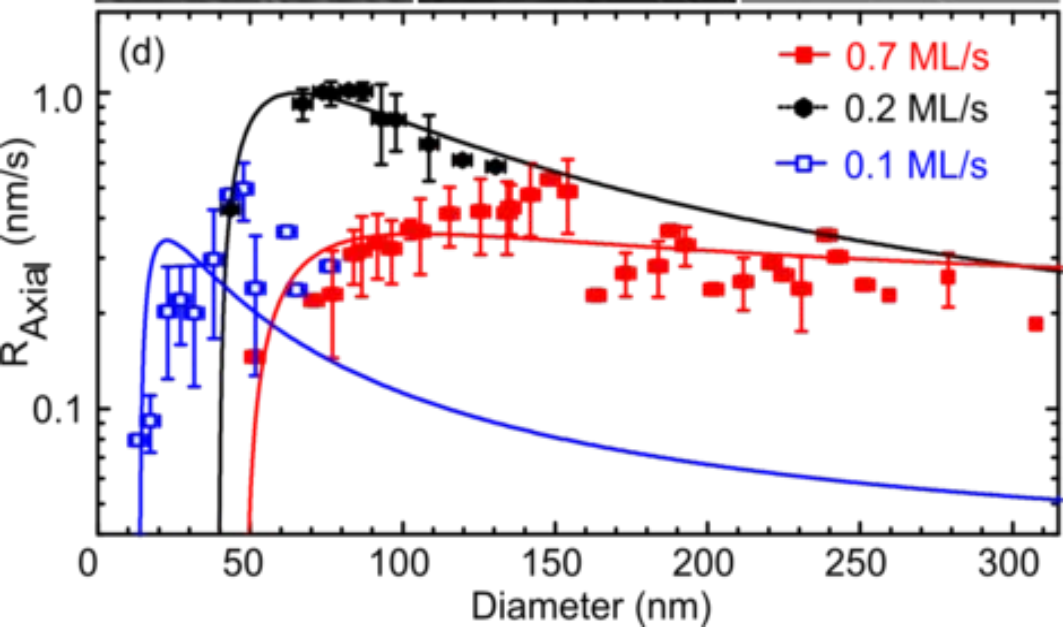

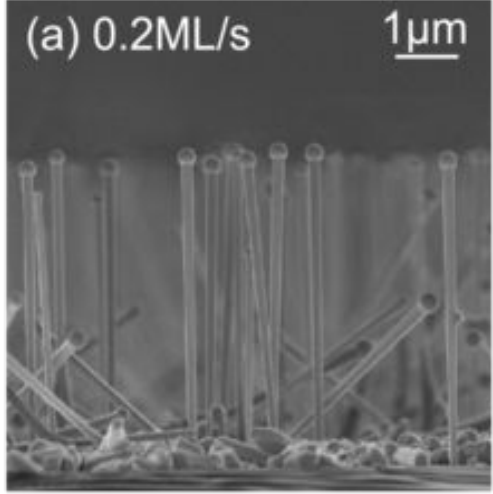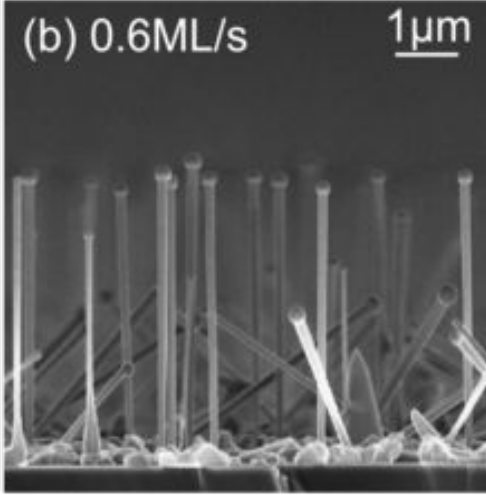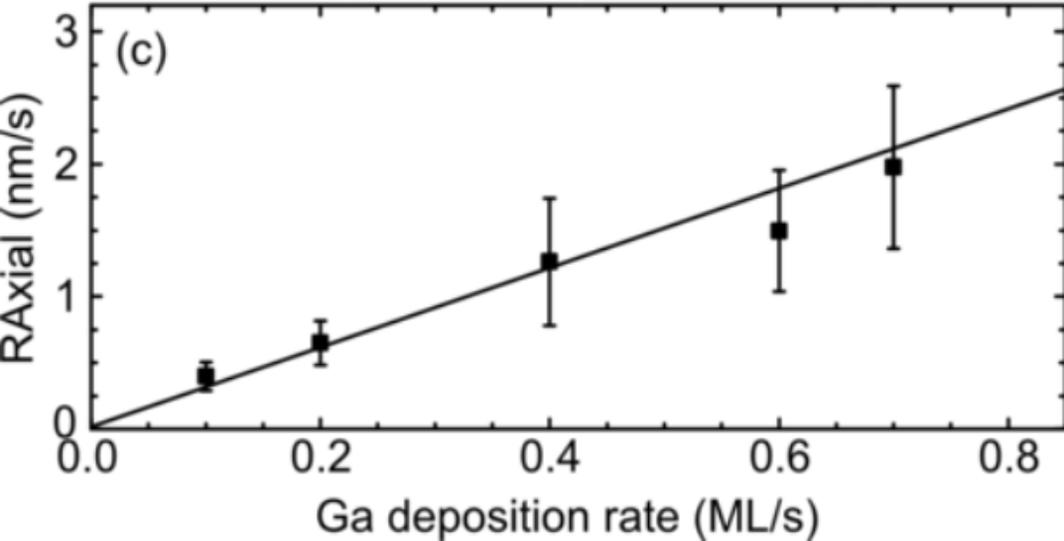